\documentclass[useAMS,usenatbib]{mn2e}

\usepackage{todonotes}
\usepackage{epsfig}
\usepackage{graphicx}
\usepackage{amsmath,latexsym,amssymb}
\DeclareGraphicsExtensions{.pdf,.png,.jpg,.eps,.ps}
\newcommand{\eq}[1]{\begin{equation}  #1 \end{equation}}

\newcommand{\eqa}[1]{\begin{align}  #1 \end{align}} % multiline equations

\newcommand{\br}[1]{\left( #1 \right)}
\newcommand{\bc}[1]{\left\{ #1 \right\}}
\newcommand{\bb}[1]{\left[ #1 \right]}

\newcommand{\nn}{\nonumber}

\newcommand{\dd}{{\rm d}}

\newcommand{\vek}[1]{\mbox{\boldmath $#1$}}
\newcommand{\ic}{{\rm i}}

\newcommand{\mpch}{\ensuremath{h^{-1}\mathrm{Mpc}}}

\newcommand{\vmax}{\ensuremath{V_{\rm max}}}

\title[Detection of correlations of fundamental plane residuals]{Detection of spatial correlations of fundamental plane residuals, and cosmological implications}
\author[B. Joachimi, S. Singh, R. Mandelbaum]{Benjamin~Joachimi,$^1$\thanks{E-mail: b.joachimi@ucl.ac.uk} Sukhdeep Singh,$^2$ and Rachel Mandelbaum$^2$\\
  $^1$Department of Physics and Astronomy, University College London, Gower Street, London WC1E 6BT, UK\\
  $^2$McWilliams Center for Cosmology, Department of Physics, Carnegie Mellon University, Pittsburgh, PA 15213, USA
}
\pagerange{\pageref{firstpage}--\pageref{lastpage}} \pubyear{2015}

\begin{document}
\label{firstpage}

\maketitle

\begin{abstract}
The fundamental plane (FP) is a widely used tool to investigate the properties of early-type
galaxies, and the tight relation between its parameters has spawned several cosmological
applications, including its use as a distance indicator for peculiar velocity surveys and as a means
to suppress intrinsic noise in cosmic size magnification measurements. Systematic trends with the
large-scale structure across the FP could cause serious biases for these cosmological probes, but may also
yield new insights into the early-type population. Here we report the first detection of spatial
correlations among offsets in galaxy size from an FP that explicitly accounts for redshift trends, using a sample
of about $95,000$ elliptical galaxies from the Sloan Digital Sky Survey. We show that these offsets
correlate with the density field out to at least $10h^{-1}$Mpc at $4\sigma$ significance in a way
that cannot be explained by systematic errors in galaxy size estimates. We propose a physical
explanation for the correlations by dividing the sample into central, satellite, and field galaxies,
identifying trends for each galaxy type separately. Central (satellite) galaxies lie on average
above (below) the FP, which we argue could be due to a higher (lower) than average mass-to-light
ratio. We fit a simple model to the correlations of FP residuals and use it to predict the impact
on peculiar velocity power spectra, finding a contamination larger than $10\,\%$ for
$k>0.04\,h/$Mpc. Moreover, cosmic magnification measurements based on an FP could be severely
contaminated over a wide range of scales by the intrinsic FP correlations.
\end{abstract}

\begin{keywords}
methods: data analysis -- cosmology: observations -- galaxies:
distances and redshifts -- galaxies: fundamental parameters -- gravitational lensing: weak -- large-scale
structure of Universe
\end{keywords}

\section{Introduction}
\label{sec:intro}

Large-scale structure cosmology has moved beyond the basic measurement of
galaxy positions used for example to infer galaxy clustering and
baryon acoustic oscillations.  There is much interest in applications
that rely on measurements of morphological properties of galaxies.  An example that
has reached a relatively mature state is weak gravitational lensing
\citep[for reviews, see][]{bartelmann01,massey10,kilbinger14}, which typically uses galaxy 
ellipticities to infer tiny but coherent shape distortions, or shear. Analogously, the sizes of galaxies can be used to
measure weak lensing magnification, which is complementary to 
shear measurements \citep[e.g.,][]{schmidt12, heavens13}, but suffers from a larger shot noise contribution due to the broad intrinsic size distribution \citep{alsing15}. This noise term can be effectively suppressed by taking advantage of the existence of a well-defined
relationship between galaxy sizes and other observables such as the
fundamental plane \citep{bertin06}. The fundamental plane (FP hereafter) corresponds to a tight relation between the size, velocity dispersion, and surface brightness of early-type galaxies \citep{1987ApJ...313...59D} and has been widely used to study the properties of ellipticals. In the context of weak lensing magnification the usefulness of the FP was recently demonstrated observationally by \citet{huff14}, using a photometric analogue.

Mapping out the peculiar velocity field \citep[e.g.,][]{springob14} is another example of the exploitation of the FP for the purposes of a cosmological probe (see \citealp{strauss95} for a review). Both applications involve measuring the observed galaxy size, and predicting a galaxy size based on the FP.  The comparison between the prediction and observation is used to measure the quantity of interest, either the size change due to 
lensing magnification, or the line-of-sight peculiar velocity, which modifies the
redshift and thus the angular diameter distance used to translate an
angular size to a physical size. 

Both of these cosmological probes make the important fundamental assumption that deviations in galaxy sizes from the FP relation do not have some underlying
correlation with the large-scale distribution of matter, other than the one induced by the cosmological signal.  Such intrinsic size
correlations are not completely unreasonable though, considering that a density dependence of FP parameters has been observed \citep[e.g.][]{barbera10}, and that 
intrinsic galaxy ellipticity correlations with the cosmic density field are a
well-established physical phenomenon \citep{troxel15}. Correlations of FP residuals with the matter density mimic the correlations between peculiar velocities or magnified galaxy sizes derived from the FP, so that the intrinsic correlations in the FP can bias the cosmological inference if not accounted for. \citet{strauss95} discussed the assumption of the universality of distance indicator relations used for peculiar velocity studies, including the FP, but concluded that, at the time, no compelling evidence for spatial or environmental variations existed.

Recently, \citet[S13 hereafter]{saulder13} used data from the Sloan Digital Sky Survey (SDSS)
to determine an FP for nearly $10^5$ early-type galaxies, with a plan of enabling measurements of the peculiar velocity field. We use this sample of unprecedented size to investigate the existence of spatial correlations of the FP residuals in the galaxy size parameter, and quantify the extent to which such correlations contaminate peculiar velocity and cosmic magnification analyses.

This paper is structured as follows: after describing in Section$\,$\ref{sec:data} the dataset that we are going to analyse, we present FP fits in Section$\,$\ref{sec:fp}, with particular focus on redshift trends. In Section$\,$\ref{sec:2pt} we summarise the estimators and error measurements for the two-point statistics we employ. Section$\,$\ref{sec:results} contains our correlation measurements, their modelling and interpretation, as well as a discussion of systematics tests. In Section$\,$\ref{sec:impact} we predict the impact of the FP residual correlations on peculiar velocity measurements and cosmic magnification, before concluding in Section$\,$\ref{sec:conclusions}.

Throughout we assume a spatially flat $\Lambda$CDM cosmology with $\Omega_{\rm
  m}=0.27$, $\Omega_{\rm b}=0.05$, $\sigma_8=0.8$, $n_{\rm s}=0.96$,
and $H_0=70\,$km/s/Mpc unless quantities are specified in terms of $h$
with $H_0=h\, 100\,$km/s/Mpc.

\section{Data}
\label{sec:data}

\begin{figure}
\centering
\includegraphics[scale=.45]{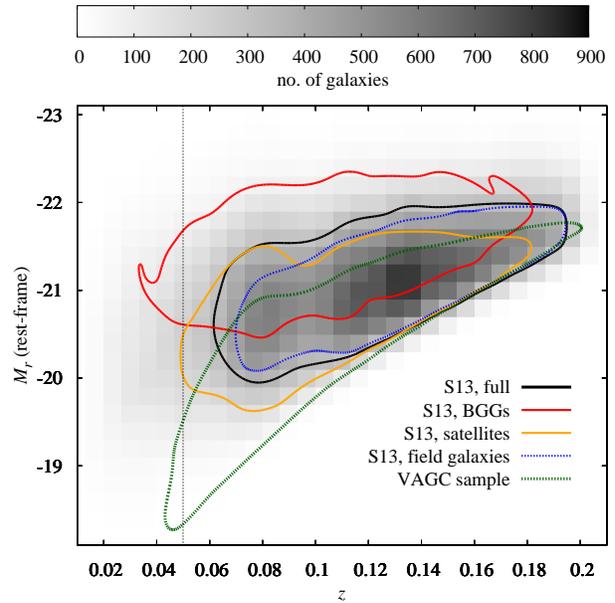}
\caption{Distributions of galaxy samples in the rest-frame $M_r$ magnitude - redshift plane. The heat map (linear scale) corresponds to the full S13 sample, with the black solid contour indicating the densest region containing $68\,\%$ of the galaxies. Corresponding contours are shown for the VAGC sample (green), the S13 brightest group galaxy (BGG) sample (red), the S13 satellite sample (yellow), and the S13 sample of field galaxies (blue). The vertical black dashed line indicates the minimum redshift cut imposed for the analysis of two-point statistics.}
\label{fig:distributions}
\end{figure} 

All data used in this paper 
originates from the Sloan Digital Sky Survey (SDSS).   The SDSS \citep{2000AJ....120.1579Y} imaged roughly $\pi$ steradians
of the sky, and followed up approximately one million of the detected
objects spectroscopically, with the relevant 
spectroscopic sample for this paper being the flux-limited Main galaxy sample \citep{strauss02}.  The imaging was carried
out by drift-scanning the sky in photometric conditions
\citep{hogg01, ivezic04}, in five bands
($ugriz$) \citep{fukugita96, smith02} using a
specially-designed wide-field camera
\citep{gunn98}.   All of
the data were processed by completely automated pipelines that detect
and measure photometric properties of objects, and astrometrically
calibrate the data \citep{2001ASPC..238..269L,pier03,tucker06}.
 The SDSS-I/II imaging
surveys were completed with a seventh data release
\citep{abazajian09}, though this work will rely as well on an
improved data reduction pipeline that was part of the eighth data
release (DR8), from SDSS-III \citep{aihara11}; and an improved
photometric calibration \citep[`ubercalibration',][]{2008ApJ...674.1217P}.

We reproduce the galaxy sample selection of S13 from SDSS DR8
as well as their derived galaxy properties. In short,
we include galaxies with spectroscopic redshifts which have been
classified as elliptical by Galaxy Zoo \citep{lintott11} with high
confidence, are better fitted by a de Vaucouleurs profile than an
exponential profile in all five passbands, and have an axis ratio $>0.3$. 
We also follow S13 in determining 
rest-frame absolute magnitudes from the de Vaucouleurs model fits, a
circularised physical galaxy radius, $R_0$, based on the de Vaucouleurs
half-light radius, an estimate of velocity dispersion, $\sigma_0$,
corrected for the effects of fixed fibre diameter, and a
dimming-corrected surface brightness, $\mu_0$, expressed in terms of\footnote{We use $\log$ to denote the logarithm to base 10, and $\ln$ for the natural logarithm.}
$\log I_0 = - 0.4 \mu_0$. After
removing a small number of duplicates and obvious outliers in the
photometry, and following the trimming of the red sequence proposed by
S13, we are left with about $95,000$ galaxies in the range $0.01 < z <
0.2$ usable for FP measurements (referred to as the S13
sample henceforth). We focus in our analysis on the $r$ and $i$ bands
as these have the best photometry as well as the tightest FP
relations.

Moreover, we also use the NYU value-added galaxy catalog\footnote{{http://sdss.physics.nyu.edu/vagc/}} \citep{blanton2005} for SDSS DR7 to construct a sample that will serve as a tracer of the matter distribution (559,028 galaxies, referred to as the VAGC sample). 
We also match the galaxies in the S13 sample with the updated SDSS DR7 group
catalog\footnote{http://gax.shao.ac.cn/data/Group.html} \citep{yang2007,yang2012} to identify the
satellites, BGGs (brightest group galaxy, treated as central galaxy) and field galaxies (galaxies in
groups of multiplicity one). We obtain 18944 (19.95\%) satellites, 18416 (19.4\%) BGGs, and 54762
(57.7\%) field galaxies in the S13 sample.  For 2812 (2.95\%) galaxies we did not find a unique match in the group
catalog, so those galaxies are not represented in figures that divide the sample based on environment.
Figure$\,$\ref{fig:distributions} shows the joint redshift and rest-frame magnitude distributions of the samples used in this work, including the environment subsamples defined above.

\section{Fundamental plane}
\label{sec:fp}

We employ a weighted linear least squares algorithm to fit FPs of the form
\eq{
\label{eq:fp}
\log R_0 = a \log \sigma_0 + b \log I_0 + c + \sum_{i=1}^{N_z} d_i z^i\;,
}
where for $N_z=0$ one recovers the standard form with free parameters
$a,b,c$. Using $N_z=0$ and \vmax\ weights, defined as the
inverse of the comoving volume in which a given galaxy would be
observable, we initially follow the analysis of S13. However, we do not
adopt the iterative clipping of S13, which is not justified as the
distribution of residuals does not feature significant outliers or
prominent tails. Consequently, the resulting root mean square residual of our fits
is about $8\,\%$ larger (see Table \ref{tab:FPfits}), while there is
good agreement in the best-fit coefficients with those in S13. Note that the analytic
expressions for the errors on the fit parameters given in S13 are only
applicable if measurement errors in $\sigma_0$ and $I_0$ were
negligible compared to those of $R_0$, and if inverse variance weights
were used, neither of which holds true in their, and our, analysis.
Instead, we estimate parameter
errors from a repeated analysis of 100 bootstrap realisations of the
underlying catalogue.

\begin{table*}
\centering
\caption{Fundamental plane (FP) fits according to Eq. \ref{eq:fp}, for
  the $r$ and $i$ bands. The fits differ by the choice of \vmax\ 
    or unit weights, and by the inclusion of an explicit
  redshift dependence (`$z$dep'). The rms deviation from the FP is
  given by $\sigma_{\rm FP}$.}
\begin{tabular}[t]{lcccccccc}
\hline\hline
model & band & $a$ & $b$ & $c$ & $d_1$ & $d_2$ & $d_3$ & $\sigma_{\rm FP}$\\
\hline
\vmax, no $z$dep    & $r$ & 1.022 $\pm$ 0.008 & -0.740 $\pm$ 0.005 & -7.79 $\pm$ 0.04 & & & & 0.100\\	

                           & $i$ & 1.050 $\pm$ 0.008 & -0.745 $\pm$ 0.005 & -7.78 $\pm$ 0.04 & & & & 0.099\\	
no \vmax, no $z$dep & $r$ & 0.925 $\pm$ 0.003 & -0.715 $\pm$ 0.001 & -7.33 $\pm$ 0.01 & & & & 0.084\\	
                           & $i$ & 0.950 $\pm$ 0.003 & -0.727 $\pm$ 0.001 & -7.37 $\pm$ 0.01 & & & & 0.082 \\
no \vmax, $z$dep    & $r$ & 0.784 $\pm$ 0.003 & -0.682 $\pm$ 0.001 & -6.86 $\pm$ 0.01 & 1.75 $\pm$ 0.11 & -9.28 $\pm$ 1.00 & 28.6 $\pm$ 2.8 & 0.077\\	
                           & $i$ & 0.816 $\pm$ 0.003 & -0.696 $\pm$ 0.001 & -6.93 $\pm$ 0.01 & 1.64 $\pm$ 0.11 & -9.50 $\pm$ 1.00 & 31.1 $\pm$ 2.9 & 0.076\\
\hline
\end{tabular}
\label{tab:FPfits}
\end{table*}

\vmax\ weights allow for the study of an ensemble of galaxies
by removing the effects of Malmquist bias.  Under the assumption that
the FP and the galaxy selection process is redshift-independent except
for the loss of faint galaxies at higher redshift due to the flux
limit, the use of \vmax\ weights essentially fills in the loss of
fainter galaxies at high redshift by proportionally upweighting them
at low redshift.  The FP derived in this way is a fair estimate of the
FP for the ensemble under the aforementioned assumptions.

\begin{figure}
\centering
\includegraphics[scale=.45]{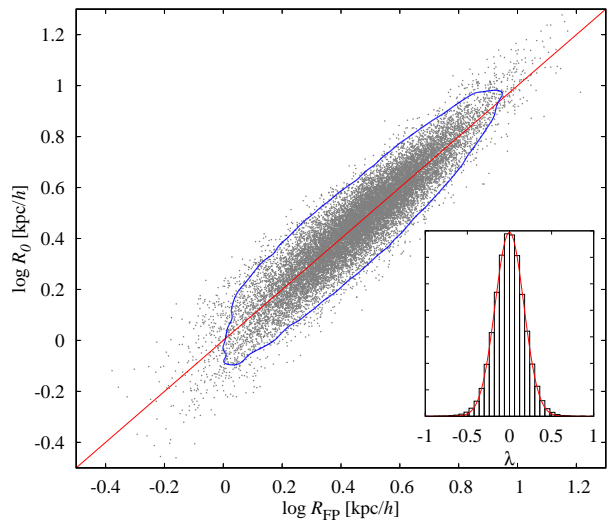}
\caption{The fundamental plane (FP) in the $r$-band used in this work (unit weights, including redshift dependence), shown for a random subset of the full S13 sample. The blue contour line encompasses $90\,\%$ of the galaxies in the most dense part of the distribution. The red line indicates a one-to-one relation. The inset shows the distribution of FP residuals $\lambda$ as defined in Eq.$\,$\ref{eq:fp_dev}, which is well described by a Gaussian (red curve).}
\label{fig:fpscatter}
\end{figure} 

\begin{figure}
\centering
\includegraphics[scale=.46]{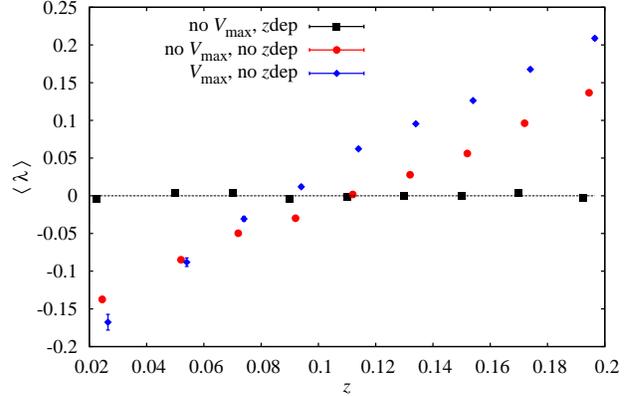}
\caption{Mean relative deviation from the FP, $\lambda$, as a function
  of redshift. Blue diamonds correspond to a FP derived with \vmax\ 
   weights and no explicit redshift dependence, as in S13. Red
  dots (black squares) show $\lambda$ for a FP with unit weights and
  without (with third-order) redshift dependence. Error bars are shown
  throughout but may be smaller than the symbols.}
\label{fig:meankappa}
\end{figure}

However, for cosmological
applications we require the FP to provide an unbiased estimate of
galaxy size for each individual galaxy. We quantify the relative
deviation of the galaxy size $R_0$ from the expectation given by the
FP, denoted $R_{\rm FP}$, via
\eq{
\label{eq:fp_dev}
\lambda \equiv \ln \frac{R_0}{R_{\rm FP}} = \frac{R_0}{R_{\rm FP}}-1 +
O\bb{\br{\frac{R_0}{R_{\rm FP}}-1}^2} \;,
}
which is an unbiased estimator of weak lensing convergence to first
order \citep[see][]{huff14}. We find $\lambda$ to be Gaussian
distributed to good approximation (see Fig.$\,$\ref{fig:fpscatter}). The average of $\lambda$ in narrow
redshift bins is shown in Fig.$\,$\ref{fig:meankappa}. While the mean
of $\lambda$ over the whole sample vanishes by construction, the standard FP with \vmax-weighting and without explicit redshift dependence shows
a strong scaling with $z$, clearly rendering it unusable
for cosmological applications, particularly the extraction of
peculiar velocities from observed redshifts.  Additionally, our
attempt to measure intrinsic size correlations using the standard FP would necessarily show a large signal even in the
absence of true intrinsic size correlations, just because correlation
functions of galaxies always pair up those that are nearby in
redshift, meaning that the average residual for the galaxies in a pair
will always have similar size and the same sign, and thus contribute a positive
correlation.

The redshift scaling of FP residuals is mainly caused by the flux limit of the SDSS Main sample which entails that we miss galaxies with small apparent size at higher redshift as they become too faint\footnote{Note that there is also an implicit lower size cut in the S13 sample, most likely driven by the requirement that they need to be reliably classified by Galaxy Zoo. However, we do not see any indications for a redshift trend in this selection effect.}. Moreover, some physically large galaxies at low redshift do not make it into the sample because they become too bright or extended to allow for a reliable photometry or morphology measurement. As a consequence, we systematically miss galaxies above the FP at low redshift and below the FP at high redshift, which leads to the tilt in the average FP residual $\lambda$ seen in Fig.$\,$\ref{fig:meankappa}. The \vmax\ weights exacerbate this trend
because they upweight small low-redshift, and thus faint, galaxies. 

Note that a physical redshift evolution of FP parameters was observed by \citet{fernandez11} in a small sample covering the redshift range from $z=0.2$ to $z=1.2$. Within the narrow redshift range of the S13 sample, such trends are dwarfed by the selection effects discussed above. However, \citet{bernardi03c} claimed to have observed weak redshift evolution in their FP, on top of selection effects caused by the apparent magnitude limit, derived from a small SDSS sample over a redshift range compatible to ours.

Switching to unit weights indeed reduces the redshift scaling of the mean residual (see
Fig.$\,$\ref{fig:meankappa}), but only by incorporating an explicit
redshift dependence with $N_z=3$ in the FP relation of Eq.$\,$(\ref{eq:fp}) does the average $\lambda$
become consistent with zero over the whole redshift range. As a
welcome by-product, the rms deviation from the FP is now $18\,\%$ less than
for the S13 results, without any manipulation of the residuals via clipping (see Table \ref{tab:FPfits}). A qualitatively similar improvement was obtained by \citet{fernandez11}. The best-fit values of $a$, $b$, and $c$ change substantially and now correspond to the FP extrapolated to $z=0$. Note that the statistical uncertainty on the three classic FP parameters remains unchanged, which suggests that they are nearly uncorrelated with $z$ as the fourth parameter in the fit. Figure$\,$\ref{fig:fpscatter} shows our optimal $r$-band FP, with unit weights and including the explicit redshift dependence, which we are going to use in the subsequent analysis.

\section{Two-point correlations}
\label{sec:2pt}

We are interested in spatial correlations of the FP residuals, $\lambda$, with each other and with the density field. To interpret these measurements, we also require the galaxy bias in order to translate a correlation with galaxy positions into a correlation with the matter distribution. We employ a suite of two-point correlation functions, measuring galaxy clustering ($\xi_{\rm gg}$), auto-correlations of $\lambda$ ($\xi_{\lambda \lambda}$), and a cross-correlation between galaxy positions and $\lambda$ ($\xi_{{\rm g}\lambda}$), with the following estimators
\eqa{	
\label{eq:defxigg}
\xi_{\rm gg} &= \frac{D D  - 2\, D R_{\rm D} + R_{\rm D} R_{\rm D}}{R_{\rm D} R_{\rm D}}\;; \\	
\label{eq:defxigl}
\xi_{{\rm g}\lambda} & =\frac{F_\lambda D - F_\lambda R_{\rm D}}{R_{\rm F} R_{\rm D}}\;; \\	
\label{eq:defxill}
\xi_{\lambda\lambda} &=\frac{F_\lambda F_\lambda}{R_{\rm F} R_{\rm F}}\;.
}
We have suppressed the arguments of the correlation functions which are calculated over a range of $(r_p,\Pi)$ bins, where $r_p$ is the comoving transverse galaxy pair separation and $\Pi$ is the  comoving line-of-sight separation. The Landy-Szalay clustering estimator \citep{Landy93} uses combinations of galaxy pair counts, denoted by $XY$ for two samples $X$ and $Y$. Here, $D$ corresponds to the density tracer samples, i.e. either S13 or VAGC, while $R_{\rm X}$ denotes sets of random points with the same selection function as sample $X$. The terms involving FP residuals are given by
\eqa{	
F_\lambda X &= \sum_{i\in F, j\in X}\lambda_i \langle j|i\rangle\;; \\
F_\lambda F_\lambda &= \sum_{i\in F, j\in F}\lambda_i \lambda_j \langle j|i\rangle\;,
}
where $\langle j|i\rangle$ is a selection function that is unity if galaxies $i$ and $j$ lie within the respective $(r_p,\Pi)$ bin and zero otherwise. Note that the subtraction of the FP residual count around random points in Eq.$\,$(\ref{eq:defxigl}) suppresses any additive systematic contribution to the correlation function $\xi_{{\rm g}\lambda}$. To reduce the shot noise contribution from randoms, we use random catalogues which are ten times larger than the real dataset when they correspond to the S13 sample, and twice larger than the real dataset when they correspond to the VAGC sample. The resulting pair counts need to be re-normalised accordingly, which we have not explicitly shown in Eqs.$\,$(\ref{eq:defxigg}) to (\ref{eq:defxill}).

We calculate the projected correlation functions via integration over $\Pi$, 
\begin{equation}\label{eqn:wab}
	w_{\rm ab} (r_p) = \int^{\Pi_{\text{max}}}_{-\Pi_{\text{max}}}\xi_{\rm ab}(r_p, \Pi)\, \mathrm{d}\Pi\;,
\end{equation}
which in practice is approximated by summation over $\Pi$ bins of width  $\Delta \Pi=10 \mpch$, with $\Pi_\text{max}=100 \mpch$. We also compare the clustering signals of galaxies above ($\lambda>0$) and below ($\lambda<0$) the FP, using the statistic
\eq{
\label{eq:deltagg}
\Delta_{\rm gg}(r_p) \equiv \frac{w_{\rm gg}(r_p, \lambda>0)}{w_{\rm gg}(r_p,\lambda<0)} -1\;.
}
A significant deviation of $\Delta_{\rm gg}$ from zero indicates a systematic variation of the clustering properties across the FP.

To get the covariance matrix for the correlation function measurement, we divide the samples into 100 jackknife 
regions of approximately equal area on the sky. Each region is about 64 square degrees in area 
($\sim8$ degrees on a side). The size of the jackknife regions constrains our ability to measure 
the covariance matrix at large physical scales. To overcome this problem, we impose a redshift cut, $z>0.05$, on our 
sample and also limit ourselves to scales $r_p \lesssim 30\mpch$ when interpreting the signals. 

While jackknifing the sample, we discard approximately 8\% of the 
S13 sample due to bright star masks and elimination of survey edges.  After masking and imposing the redshift cut, we are left with 80,550 galaxies in the S13 sample 
and 423,053 galaxies in the VAGC sample that we use in the final measurements presented in the following section. While these masks are uncorrelated with large-scale structure, they could in principle degrade the completeness and purity of the environment classification by spuriously removing whole galaxy groups or group members from the sample. \citet{yang2007} correct for the impact of survey edges, and our masking actually alleviates this effect further by removing the survey regions with the most complex geometries. Bright star masks only account for a few per cent of the total area masked, so that their impact is negligible.

We find strong correlations between the correlation function measurements in different $r_p$ bins on all scales considered, in particular for $w_{\rm gg}$, so that the full jackknife covariances are used throughout. When inverting these covariances for likelihood analysis, we account for the noise due to the finite number of jackknife samples by employing an approximately unbiased estimator for the inverse covariance, following \citet{kaufman67}.

We carry out the correlation function measurements separately in the $r$ and $i$ bands.
The two-point functions in the two bands are highly correlated, so there
is not much extra cosmological information when adding another passband,
but requiring consistency of the results is a useful sanity check.

\section{Signals and modelling}
\label{sec:results}

\subsection{Clustering}

We measure the galaxy clustering signal of our density tracer samples with the primary goal of obtaining an effective linear and deterministic galaxy bias, $b_{\rm g}$. Closely following the formalism of \citet{baldauf10}, we construct the following model for the projected correlation function,
\eqa{
\label{eq:ggmodel}
w_{\rm gg}(r_p) & = 2\, b_{\rm g}^2 \int_0^\infty \!\!\!\! \dd z\, {\cal W}(z) \sum_{l=0}^2 \alpha_{2 l} \bb{\frac{f(z)}{b_{\rm g}}} \int_0^{\Pi_{\rm max}} \!\!\!\!\!\!\!\!\!\! \dd \chi  \\ \nn
 \times\; & \xi_{{\rm \delta \delta},2 l} \br{\sqrt{\chi^2+r_p^2},z} L_{2 l} \br{\frac{\chi}{\sqrt{\chi^2+r_p^2}}} +\; C_{\rm IC} \;,
}
where we introduced the growth rate $f(z)=\dd \bb{\ln D(z)}/\dd \bb{\ln a}$, with $D(z)$ the growth factor and $a=1/(1+z)$ the cosmic scale factor. The $L_{2 l}$ denote Legendre polynomials, and the $\alpha_{2 l}$ are polynomials in the variable $\beta=f(z)/b_{\rm g}$ whose explicit form is given in Eqs. (48) to (50) of \citet{baldauf10}. We relate the multipoles of the matter correlation function to the matter power spectrum, $P_\delta$, as follows,
\eq{
\xi_{{\rm \delta \delta},2 l}(r,z) = \frac{(-1)^l}{2 \pi^2} \int_0^\infty \dd k\, k^2 j_{2 l}(k r)\; P_\delta(k,z)\;,
}
where the $j_{2 l}$ are spherical Bessel functions. Throughout, we employ the full non-linear matter power spectrum as a means to phenomenologically extend our models into the mildly non-linear regime, using the updated \texttt{halofit} non-linear fit \citep{takahashi12} and the transfer function fit by \citet{eisenstein98}.

We average the correlation function model over redshift with the weight
\eq{
\label{eq:zaverage}
{\cal W}(z) = \frac{p_{\rm a}(z)\; p_{\rm b}(z)}{\chi^2(z)\; \chi'(z)}\; \bb{\int \dd z\; \frac{p_{\rm a}(z)\; p_{\rm b}(z)}{\chi^2(z)\; \chi'(z)}}^{-1}\;,
}
as motivated and derived in \citet{mandelbaum11}. The prime denotes the derivative of comoving distance with respect to redshift. We have written the weight more generally for the cross-correlation of two samples a and b, with redshift probability distributions $p_{\rm a}(z)$ and $p_{\rm b}(z)$, respectively. We follow \citet{roche99} in estimating the integral constraint, $C_{\rm IC}$, via the random pair counts. Since the projected correlation function is dominated by contributions of pairs with small line-of-sight separation, we find it is a fair approximation to calculate $C_{\rm IC}$ using the projected matter correlation function, $w_{\delta \delta}(r_p)$, in combination with the random pair counts stacked along the line of sight. Moreover, the contributions by multipoles beyond $l=0$ in Eq.$\,$(\ref{eq:ggmodel}) are small and hence can safely be neglected in the calculation of the integral constraint. Consequently, we compute
\eq{
\label{eq:ic}
C_{\rm IC} \approx \frac{\sum_{r_p} \bb{R_{\rm D} R_{\rm D}}(r_p)\; w_{\delta \delta}(r_p) }{\sum_{r_p} \bb{R_{\rm D} R_{\rm D}}(r_p)}\;,
}
where $w_{\delta \delta}$ is given by the monopole of Eq.$\,$(\ref{eq:ggmodel}), divided by $b_{\rm g}^2$.

Using random pair counts over the range $r_p=\bb{0.2;200}\,{\rm Mpc}/h$, we obtain $C_{\rm IC} = 0.29\,{\rm Mpc}/h$ for the NYU VAGC sample and $C_{\rm IC} = 0.26\,{\rm Mpc}/h$ for the S13 sample. The values for $C_{\rm IC}$ are at least an order of magnitude smaller than $w_{\rm gg}$ at the largest scales we consider. We then fit the model of Eq.$\,$(\ref{eq:ggmodel}) with $b_{\rm g}$ as a single free parameter, using four bins over the range $r_p=\bb{5.6;30}\,{\rm Mpc}/h$. This results in a reduced $\chi^2$ close to unity, with best-fit parameters $b_{\rm g}=1.18 \pm 0.09$ for the NYU VAGC sample and $b_{\rm g}=1.37 \pm 0.09$ for the slightly more luminous S13 sample (errors are $1\sigma$), in good agreement with previous results (see e.g. \citealp{zehavi11} who also worked on SDSS Main samples and used a similar fit range).

\subsection{FP residual size correlations}

\begin{figure}
\centering
\includegraphics[scale=.48]{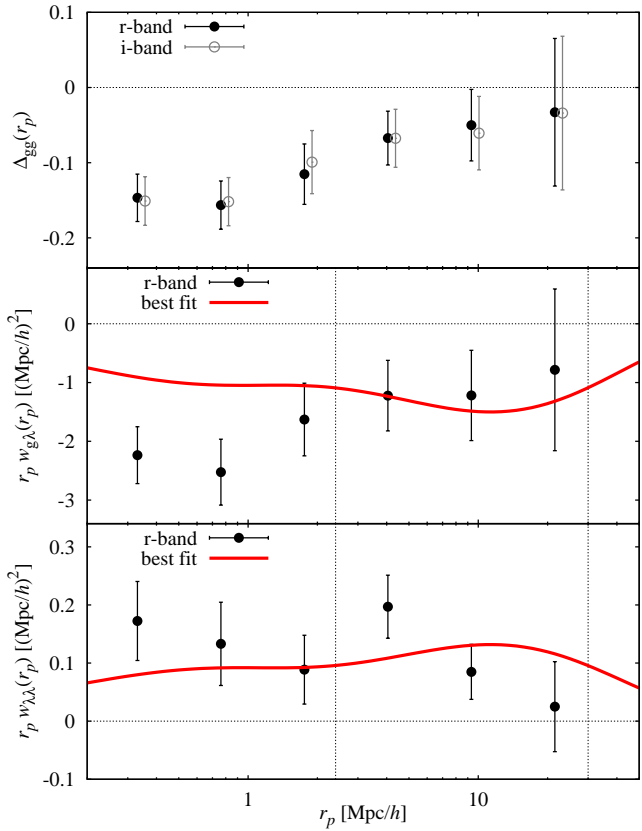}
\caption{\textit{Top}: relative difference between the clustering
  correlation of galaxies above the fundamental plane ($\lambda>0$) and below the fundamental plane ($\lambda<0$), as a function of comoving transverse galaxy
  separation $r_p$. Results for the $r$-band (black dots)
  and $i$-band (grey circles) are very
  similar. \textit{Centre/bottom}: galaxy density-size correlation
  function, $w_{{\rm g} \lambda}$ (centre), and galaxy size-size
  correlation function, $w_{\lambda \lambda}$ (bottom). The best-fit
  model for each individual signal is shown as the red solid line. 
  The vertical lines delimit the $r_p$ range used for the fit. Note that the correlation functions have been rescaled by a factor $r_p$. Throughout, the S13 sample was used as the density tracer.}
\label{fig:2pt}
\end{figure} 

In Fig.$\,$\ref{fig:2pt} we show results for the spatial correlations of FP residuals $\lambda$ using the statistics $\Delta_{\rm gg}$, $w_{{\rm g} \lambda}$, and $w_{\lambda \lambda}$ as defined in Sect.$\,$\ref{sec:2pt}. The corresponding significance of the signals, each measured over the range $r_p=\bb{0.2;30}\,{\rm Mpc}/h$ and determined using the full covariance matrix of the data points, is given in Table \ref{tab:signalfits}. We observe a significant negative correlation in both $\Delta_{\rm gg}$ and $w_{{\rm g} \lambda}$, plus a tentative positive signal in $w_{\lambda \lambda}$, with no dependence on the passband in which the FP was determined. Using the S13 sample also for the galaxy number densities, produces marginally stronger detections because it traces slightly more massive structures on average (see Fig.$\,$\ref{fig:distributions}). A negative $w_{{\rm g} \lambda}$ correlation implies that galaxies below the FP, i.e. with negative $\lambda$, are preferentially located in environments with higher density. This is in line with the negative clustering ratio, $\Delta_{\rm gg}$, which means that the galaxies below the FP are more strongly clustered, again pointing towards a high-density environment for these objects. 

\begin{table}
\centering
\caption{Significance of correlation signals in multiples of the width $\sigma$ of a Gaussian distribution, measured over the range $r_p=\bb{0.2;30}\,{\rm Mpc}/h$. Columns indicate whether the S13 or VAGC samples were used as the density tracer, and whether the $r$ or $i$ bands were used to determine the fundamental plane. The
  bottom half of the table lists systematics tests, re-computing the
  statistic with reshuffled sky coordinates (`rand. pos.') or using
  only galaxy pairs with line-of-sight separation in the range $\Pi
  \in \bb{200;400}\,{\rm Mpc}/h$ (`large $\Pi$').}
\begin{tabular}[t]{lcccc}
\hline\hline
signal & \multicolumn{4}{c}{significance ($x \sigma$)} \\
 & S13, $r$ & S13, $i$ & VAGC, $r$ & VAGC, $i$ \\
\hline
$w_{\rm g \lambda}$ & 3.80 & 4.18 & 3.17 & 3.07\\
$w_{\rm \lambda \lambda}$ & 2.71 & 2.36 & 2.71 & 2.36\\
$\Delta_{\rm gg}$ & 4.28 & 4.40 & 3.35 & 3.92\\
\hline
$w_{\rm gg}$, large $\Pi$ & 1.51 & 1.25 & 1.02 & 1.02\\
$w_{\rm g \lambda}$, large $\Pi$ & 0.34 & 0.52 & 0.82 & 0.95\\
$w_{\rm \lambda \lambda}$, large $\Pi$ & 0.67 & 0.83 & 0.67 & 0.83\\
$w_{\rm g \lambda}$, rand. pos. & 0.09 & 0.06 & 0.43 & 0.27\\
$w_{\rm \lambda \lambda}$, rand. pos. & 0.23 & 0.37 & 0.23 & 0.37\\
\hline
\end{tabular}
\label{tab:signalfits}
\end{table}

\begin{figure}
\centering
\includegraphics[scale=.46]{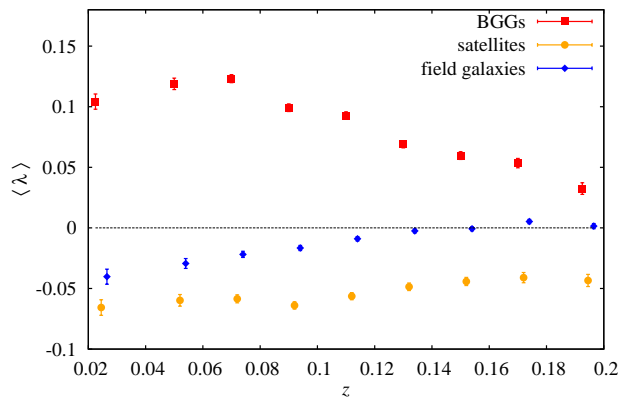}
\caption{Mean relative deviation from the fundamental plane, $\lambda$, as a function
  of redshift for subsamples of S13 defined according to environment. The colour coding is the same as in Fig.$\,$\ref{fig:distributions}. Note that this was obtained for the $r$-band fundamental plane fit without \vmax\ weights and with redshift dependence included.}
\label{fig:kappa_environment}
\end{figure}

To further elucidate this environment dependence, we compute the mean FP residuals for the different environment subsamples introduced in Sect.$\,$\ref{sec:data}. As is evident from Fig.$\,$\ref{fig:kappa_environment}, the brightest galaxies in groups (BGGs) tend to lie well above the FP, while satellite galaxies have preferentially negative $\lambda$, with both trends showing only mild redshift evolution over the range we can probe. Isolated (\lq field\rq) galaxies constitute by far the largest subsample (see Section \ref{sec:data}). Their mean $\lambda$ is driven towards slightly negative values, in particular at low redshift, to compensate for the strongly positive $\lambda$ of BGGs because, by construction, our FP forces the mean $\lambda$ of the full sample to zero in every redshift bin. Since field galaxies and satellites dominate the contributions to $w_{\rm g \lambda}$ because of their numbers, the resulting correlation is negative.

The trends in Fig.$\,$\ref{fig:kappa_environment} can also explain the result of negative $\Delta_{\rm gg}$. Since a halo can by definition host at most one BGG but arbitrarily many satellites, the latter are expected to reside on average in more massive haloes which therefore are upweighted in correlations of satellites. Indeed, the haloes of BGGs in the S13 sample have an average number of 4.4 members while for satellites this number is 16.7. Consequently, we expect the satellite subsample to have higher galaxy bias than the BGG subsample. Since they are of approximately equal size, galaxies below the FP should cluster more strongly than those above the FP, leading to $\Delta_{\rm gg}<0$. Field galaxies do not drive the effect because they are more weakly clustered, especially on the mildly non-linear scales for which we detect a signal, and, besides, are more symmetrically distributed above and below the FP.

In summary, the spatial correlations of FP residuals we observe appear to be linked to a dependence of the FP on the galaxies\rq\ environment. \citet{bernardi03c} and \citet{barbera10} both investigated the dependence of FP parameters on estimates of local density, using large SDSS samples at low redshift. Their results can be re-interpreted as the FP residual, $\lambda$, smoothly increasing with density, in line with our findings for BGGs. Analogous trends have also been observed for the central galaxies of more massive systems \citep[e.g.][]{bernardi07,linden07}.

However, none of these works considered satellite galaxies separately. It is suggestive to relate the consistently negative $\lambda$ for satellite galaxies, as seen in Fig.$\,$\ref{fig:kappa_environment}, to systematic trends in galaxy size at fixed mass. Evidence for such trends is either weak \citep{2010ApJ...709..512R,2010MNRAS.402..282M} or opposite to the effects we see \citep{2012ApJ...750...93P}, but note that these works used far smaller samples. Moreover, trends in $\lambda$ could equally well be caused by a density dependence of the velocity dispersion, $\sigma_0$, or the surface brightness, $I_0$. \citet{bernardi03a} found that, in high-density regions, $\sigma_0$ and $R_0$ increase while $I_0$ decreases, although all dependencies were weak (see also \citealp{2012MNRAS.419.3018C} for a similar result on the $R_0$ dependence on density as measured in DEEP2/3).

Systematic deviations from the FP have also been interpreted as being due to variations in the mass-to-light ratio, $M/L$, of galaxies \citep{graves10}. \citet{barbera10} accordingly concluded from their FP analysis that galaxies in high-density environments have higher $M/L$ on average. We can confirm this trend for BGGs and additionally infer that satellite galaxies have lower $M/L$ than corresponding galaxies in the field, in agreement with the picture of mass stripping of galactic subhaloes falling into larger group and cluster haloes.

Note that we have repeated the FP fit of Section$\,$\ref{sec:fp} with either a local galaxy number density estimate or the distance to the nearest neighbour galaxy as an additional variable, but have found neither a  change in the best-fit FP parameters nor a decrease in the scatter of the FP residuals. This implies that these two variables are not sufficiently good indicators of the environment dependence of the FP that leads to the spatial correlations observed in this work.

We propose a simplistic model for the spatial correlations of the FP residuals, assuming that $\lambda$ is proportional to the matter density contrast, $\delta$,
\eq{
\label{eq:lambdamodel}
\lambda(\vek{x}) := B\, \delta(\vek{x})\;.
}
There is currently no clear motivation for a physical ansatz; this choice is just one of the simplest possible. \citet{alsing15} used an analogous approach for modelling galaxy size residuals \emph{not} related to an FP (see also \citealp{ciarlariello14} for a more sophisticated ansatz). Using Eq.$\,$(\ref{eq:lambdamodel}), the projected correlation functions read
\eqa{
\label{eq:2ptmodels}
w_{{\rm g} \lambda}(r_p) &= b_{\rm g}\, B\, w_{\delta \delta}(r_p)\;; \\
w_{\lambda \lambda}(r_p) &= B^2\, w_{\delta \delta}(r_p)\;, 
}
with
\eq{
\label{eq:wdd}
w_{\delta \delta}(r_p) = \int_0^\infty \dd z\, {\cal W}(z) \int_0^\infty \frac{\dd k_\perp\, k_\perp}{2\,\pi}\, J_0(k_\perp r_p)\, P_{\delta} (k_\perp,z)\;, 
}
where $J_0$ is the Bessel function of the first kind of order zero. Here, we have assumed that the chosen $\Pi_{\rm max}$ is sufficiently large that the line-of-sight integration in the model can be extended to infinity. Furthermore, we have neglected redshift-space distortions. Note that the redshift distributions of the S13 and VAGC samples are sufficiently similar that we expect negligible differences between the effective galaxy bias as measured from the VAGC clustering signal and the galaxy bias entering the cross-correlation $w_{{\rm g} \lambda}$ with the VAGC sample as the density tracer.

We fit this model over the range $r_p=\bb{2.4;30}\,{\rm Mpc}/h$ separately to the correlation functions with the amplitude $B$ as the free parameter. Results are given in Table \ref{tab:modelfits}, and the best-fit model in each case is also over-plotted in Fig.$\,$\ref{fig:2pt}. We refrain from including smaller scales as the assumption of an effective linear galaxy bias, as well as the model of Eq.$\,$(\ref{eq:lambdamodel}), are likely to break down. This implies that the model parameter constraints are weaker than the original detection significance down to sub-megaparsec scales would suggest.

The constraints for $w_{\rm g \lambda}$ are independent of the passband and density tracer and yield $B \approx -0.01$ with about a $50\,\%$ error ($1\sigma$; note the error on galaxy bias has been propagated but is negligible). The reduced $\chi^2$ is quite small in this case, which may indicate that the jackknife covariances provide very conservative errors in the $r_p$ range of the fit. In contrast, the model yields a bad fit to $w_{\rm \lambda \lambda}$ and prefers an amplitude $B$ that is substantially larger (note this statistic is not sensitive to the sign of $B$). We speculate that this could be a sign of a failure of our model. Higher-order contributions from a non-linear relation between $\lambda$ and $\delta$ or a density weighting (see \citealp{hirata04} for an analogous calculation in the context of intrinsic galaxy alignments) could have more impact on $w_{\rm \lambda \lambda}$ than on $w_{\rm g \lambda}$ and boost in particular the smallest scales included in the fit range.

\begin{table}
\centering
\caption{Constraints on the intrinsic size correlation amplitude $B$, using the range $r_p=\bb{2.4;30}\,{\rm Mpc}/h$. The constraints on $w_{\rm g
    \lambda}$ were obtained with both the S13 sample (S) and the VAGC sample (V) as the density
  tracer. Note that $w_{\rm \lambda \lambda}$ can only constrain the absolute value of $B$.}
\begin{tabular}[t]{lcccc}
\hline\hline
signals & \multicolumn{2}{c}{$r$-band} & \multicolumn{2}{c}{$i$-band}\\
 & $B$ (1$\sigma$) & $\chi^2_{\rm red}$ & $B$ (1$\sigma$) & $\chi^2_{\rm red}$\\
\hline
$w_{\rm g \lambda}$ (S)  & $-0.012 \pm 0.005$ & 0.08 & $-0.010 \pm 0.005 $ & 0.02 \\
$w_{\rm g \lambda}$ (V) & $-0.009 \pm 0.006$ & 0.18 & $-0.009 \pm 0.005 $ & 0.08 \\
$w_{\rm \lambda \lambda}$   & $ \pm 0.038 \pm 0.006$ & 2.81 & $ \pm 0.035 \pm 0.006$ & 2.43 \\
\hline
\end{tabular}
\label{tab:modelfits}
\end{table}

\subsection{Systematics tests}
\label{sec:systematics}

We also carry out several tests for systematic errors.  The first test
is to measure $w_{g\lambda}$ and
$w_{\lambda\lambda}$ with the line-of-sight integral in Eq.~\eqref{eqn:wab} ranging from
  $200$ to $400\,$Mpc/$h$ instead, considering both positive and negative values of $\Pi$ (referred to in plots
  and tables as the `large  $\Pi$' test).  On these large scales, true intrinsic FP residual
  correlations should be negligible, so this measurement can reveal
  whether there may be spurious signals due to image processing
  systematics, such as misestimation of the sky level around bright
  galaxies (which affects the flux and size measurements of nearby
  galaxies out to 100\arcsec; see, e.g., \citealt{aihara11} for
  estimates of the severity of the effect in DR8 data).  Note
  that for very large $\Pi$ we expect non-negligible measurements of
  $w_{g\lambda}$ and $w_{\lambda\lambda}$ due to lensing
  magnification, but the separations used are small enough that we do not
  expect any detection of this effect. Table \ref{tab:signalfits} lists the significances of correlation signals in the large $\Pi$ statistic, measured over the range $r_p=\bb{0.2;30}\,{\rm Mpc}/h$. The results for all two-point statistics are comfortably consistent with zero. Note that the measurements for different passbands and density tracers are highly correlated, so that similar outcomes for the detection significances are expected across the columns of Table \ref{tab:signalfits}.
  
Another systematics test involves permuting the sky coordinates of the galaxies, i.e. randomising the (RA,DEC) pairs in relation to the redshift and FP residual. This test (referred to as `rand. pos.') is particularly sensitive to a non-zero mean of $\lambda$ as a function of redshift; see the issues raised in Section~\ref{sec:fp}. Neither $w_{g\lambda}$ nor $w_{\lambda\lambda}$ show any signs of significant signals in this statistic for our final unit-weight and redshift-dependent FP (see Table \ref{tab:signalfits}), although both were clearly non-zero for the \lq classic\rq\ choice of FP discussed in Section \ref{sec:fp}. Note that we observe a low-level negative correlation of $\lambda$ with random points, which is constant as a function of $r_p$. This systematic remains below the signal in absolute value on all scales and is subtracted off automatically by the estimator of Eq.$\,$(\ref{eq:defxigl}).

When computing the model for our correlation functions (see Eq.~\ref{eq:wdd}), we assume $\Pi_{\text{max}} \rightarrow \infty$ in Eq.~(\ref{eqn:wab}). This is justified since our choice of $\Pi_{\text{max}}=100\mpch\ $ when computing the correlation function from the data is large enough to capture almost all of the information. Still, to test for the choice of  $\Pi_\text{max}$ and $\Delta \Pi$, we repeat the analysis with a) $\Pi_\text{max}=50 \mpch$ and $\Delta \Pi=5 \mpch$, and b) $\Pi_\text{max}=200 \mpch$ and $\Delta \Pi=20 \mpch$. Using the $r$-band and the S13 sample as density tracer, we find $B = -0.011 \pm 0.005$ for case a) and $B = -0.011 \pm 0.006$ for case b), fully consistent with the results for the default choice of $\Pi_\text{max}$ and $\Delta \Pi$ (see Table \ref{tab:modelfits}). Hence, conclusions are not significantly affected by our particular choice of line-of-sight binning.

It is known that the standard SDSS pipeline has issues with the estimation of sky background in crowded regions, and especially for large objects \citep[e.g.,][]{2006ApJS..162...38A,aihara11}. This could induce spurious environment-dependent galaxy size measurements and thus affect the spatial correlations we wish to detect. \citet{hyde09a} found that the SDSS pipeline parameters underestimate the total flux and size for large galaxies when compared to their own data reduction. However, note that these authors worked on DR6 while we use the galaxy parameters from DR8 for which the treatment of large galaxies was substantially improved \citep{aihara11}. None the less, we employ the magnitude and radius corrections in Eqs. (4) and (5) of \citet{hyde09a} to re-fit an FP and re-measure the $w_{\rm g \lambda}$ and $w_{\rm \lambda \lambda}$ statistics, as a conservative estimate of the influence of sky subtraction effects. We find that these corrections modify the correlation functions by much less than $1\sigma$ on all scales considered, so our results are expected to be robust against such systematics.
Finally, the consistency of measurements in the $r$ and $i$ bands also underlines that our analysis is not significantly affected by issues with any photometric measurements that affect these passbands differently.

\section{Impact on cosmology}
\label{sec:impact}

The small scatter of galaxy radii around the fundamental plane has been exploited in several applications of large-scale structure cosmology. In the following we will estimate the bias incurred by two such applications if the spatial correlations between FP residuals detected in this work are not accounted for.

\subsection{Peculiar velocity power spectrum}
\label{sec:vpec}

The peculiar velocities of galaxies probe the slopes of gravitational potentials and thus can be used to constrain the growth rate of structure, $f(z)$. If an independent distance indicator is available, the observed redshift, $z$, can be split into contributions by the smooth Hubble expansion, denoted by $z_{\rm H}$, and by the line-of-sight component of the peculiar velocity, $u$. The relation reads \citep{harrison74}
\eq{
\label{eq:zrelation}
\br{1+z} = \br{1 + z_{\rm H}}\; \br{1+\frac{u}{c}}\;.
}
The FP is a widely used distance indicator in peculiar velocity studies. It is assumed that any systematic offset in radius from the FP is caused by the peculiar velocity of the galaxy, where $R_0$ is determined from the observed redshift, and $R_{\rm FP}$ from $z_{\rm H}$. The translation from galaxy radius to redshift is done via the angular diameter distance, $D_{\rm A}(z)=\chi(z)/(1+z)$, where $\chi$ denotes comoving distance; see \citet{johnson14} for details.

The spatially correlated contribution to $\lambda$, as defined in Equation (\ref{eq:fp_dev}), adds a systematic shift, $\Delta u$, to the inferred line-of-sight component of the peculiar velocity, $u_{\rm obs}$. We linearise the response of $u_{\rm obs}$ to $\lambda$ and write
\eq{
\label{eq:uexpansion}
u_{\rm obs} = u + \Delta u \approx u + \left. \frac{\dd u}{\dd \lambda} \right|_{\lambda=0}\, \lambda \equiv u + T(z)\, \lambda\;,
}
where $T(z)$ encapsulates the response. Using Equation (\ref{eq:zrelation}), and linking $z_{\rm H}$ via $D_A(z_{\rm H})$ and $R_{\rm FP}$ to $\lambda$, we obtain the expression
\eqa{
\label{eq:transfer}
T(z_{\rm H}) & = - c\, \frac{1+z}{1+z_{\rm H}} \bb{ (1+z_{\rm H})\; \frac{\chi'(z_{\rm H})}{\chi(z_{\rm H})} - 1 }^{-1}\\ \nn
& \approx - c \bb{ (1+z_{\rm H})\; \frac{\chi'(z_{\rm H})}{\chi(z_{\rm H})} - 1 }^{-1}\;,
}
where $\chi'$ denotes the derivatives of comoving distance with respect to redshift. The first equality agrees with the result of \citet[Eqs.$\,7-9$]{johnson14}. Note that our expression lacks a factor $\ln 10$ since we employ the natural logarithm in $\lambda$ rather than the decadic one, and that we find a different sign in the Jacobian of the mapping between $\lambda$ and $\log \bb{ D_A(z)/D_A(z_{\rm H})}$, in agreement with \citet{springob14}. The approximation in the second equality is equivalent to assuming $u/c \ll 1$, which is fair in the context of this forecast.

\begin{figure}
\centering
\includegraphics[scale=.45]{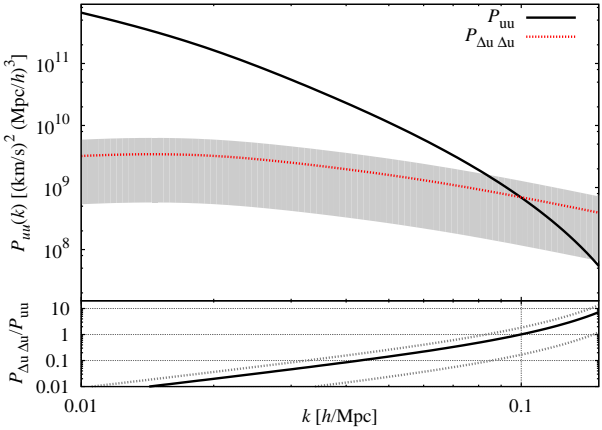}
\caption{Predicted contamination by intrinsic fundamental plane residual correlations of the
  peculiar velocity auto-correlation power spectrum, $P_{uu}(k,\mu=1)$, at
  $z=0.1$. In the top panel the cosmological signal is
  given by the black solid line and the systematic contribution by
  the red dotted line. The best-fit value for $B$ from the $w_{\rm g
    \lambda}$ fit ($r$-band, S13 sample as density tracer) has been propagated. Its propagated
  $1\sigma$ error results in the error envelope shown in grey. The bottom panel gives the ratio of
  the systematic over cosmological signal, with $1\sigma$ uncertainty indicated by the grey dotted lines.
}
\label{fig:forecasts_vpec}
\end{figure} 

Using linear theory, the peculiar velocity component $u$ can be linked to the matter density contrast, $\delta$ \citep[e.g.][]{hamilton98}. An analogous link can be established for the shift $\Delta u$ by combining Equations (\ref{eq:uexpansion}) and (\ref{eq:lambdamodel}), which results in
\eq{
\label{eq:ufourier}
\tilde{u}_{\rm obs}(\vek{k}) = \bb{ - \ic H(z) (1+z)^{-1} f(z) \frac{\mu}{k} \ + T(z) B} \tilde{\delta}(\vek{k})\;,
}
where tildes mark Fourier transformed quantities, and $\mu$ is the cosine of the angle between the wave vector $\vek{k}$ and the line-of-sight direction (see also \citealp{koda14} for a motivation of the first term in Eq.$\,$\ref{eq:ufourier}). Here, $H(z)$ is the Hubble parameter. The \lq observed\rq\ velocity power spectrum (containing the combined effects of peculiar velocity and intrinsic size correlations) then reads
\eq{
\label{eq:velocityps}
P_{uu}^{\rm obs}(k,\mu,z) = \bc{ \frac{H^2(z) f^2(z)\, \mu^2}{(1+z)^2\, k^2} + T^2(z) B^2 }\, P_{\delta \delta}(k,z)\;.
}
The first term corresponds to the standard velocity power spectrum \citep[see e.g.][]{koda14}, while the second term marks a purely additive systematic contribution by intrinsic FP residual correlations. In Fig.$\,$\ref{fig:forecasts_vpec} we plot the two terms and the relative strength of the contamination, for $\mu=1$ and including the empirical damping term proposed by \citet{koda14}. We have propagated the best-fit value and error of $B$ for the fit to $w_{{\rm g}\lambda}$ in the $r$-band and using the S13 sample as the density tracer. Intrinsic FP residual correlations could constitute a significant systematic effect for the peculiar velocity power spectrum above the $10\,\%$ level for $k > 0.04\,h/$Mpc and may dominate the signal according to our best-fit model beyond $k=0.1\,h/$Mpc. While the statistical uncertainty on this statement is still large and a null signal for the contaminant is not rejected with high confidence (due to the limited model fitting range), this finding suggests that any cosmological analysis based on peculiar velocity measurements from the FP should at least test for the presence of this systematic.

\citet{johnson14} performed a cosmological peculiar velocity analysis using the 6dFGRS, measuring the velocity power spectrum out to $k=0.15\,h/$Mpc. They found no indication for an excess signal on small scales, nor any discrepancies with the signals as derived from a sample of type Ia supernovae. However, the statistical error bars are sufficiently large to accommodate a systematic at the few tens of percent level. Note that, since the authors recast their measurements into apparent magnitude correlations, which is beneficial for statistical reasons, they may be prone to yet another systematic as it seems plausible that galaxies also exhibit correlations of their intrinsic fluxes.

There are several avenues to mitigate any systematics arising from intrinsic size correlations. Apart from contrasting measurements with those using other distance indicators, one can check the dependence of the velocity power spectrum on $\mu$. Since the systematic contribution derives from an effect in the plane of the sky, it does not depend on $\mu$ and could thus be filtered out as a constant offset. Moreover, cross-correlations between galaxy position and peculiar velocity should be helpful in calibrating out intrinsic size correlations. In this statistic the galaxy-velocity power is purely imaginary \citep{koda14}, while intrinsic size correlations generate a real contribution, provided $\lambda$ can be modelled as a local function of the matter density contrast.

\subsection{Weak lensing magnification}
\label{sec:mag}

Probing weak gravitational lensing magnification via the spatial correlations of apparent galaxy sizes is regarded as a promising complement to the established gravitational shear measurements \citep{heavens13}. The signal-to-noise ratio of such measurements is limited by the intrinsic scatter of galaxy sizes, which is substantially larger than for the corresponding measure of galaxy ellipticity \citep{alsing15}. Therefore it could be desirable to instead use the much smaller scatter of galaxy sizes around an FP, even if this could only be obtained for a much smaller early-type galaxy sample (see \citealp{huff14} for a recent application). Intrinsic correlations of FP residuals constitute a systematic for this type of weak lensing measurement that is closely analogous to the issue of intrinsic galaxy alignments in gravitational shear statistics \citep{troxel15}.

We assess the impact of intrinsic size correlations on the convergence power spectrum,
\eq{
\label{eq:limber_lensing}
C^{(ij)}_{\kappa \kappa}(\ell) = \int^{\chi_{\rm hor}}_0 \dd \chi\;
\frac{q^{(i)}(\chi)\; q^{(j)}(\chi)}{\chi^2}\; P_{\delta \delta}
\br{\frac{\ell}{\chi},\chi}\;, 
}
which is a statistic that can be derived from estimators of the form given in Equation (\ref{eq:fp_dev}). The integration limit $\chi_{\rm hor}$ is the comoving distance to the horizon, and the lensing kernel, $q$, is given by
\eq{
\label{eq:weightlensing}
q^{(i)}(\chi) = \frac{3 H_0^2 \Omega_{\rm m}}{2\, c^2}
\frac{\chi}{a(\chi)} \int_{\chi}^{\chi_{\rm hor}} \dd \chi_{\rm s}\;
p^{(i)}(\chi_{\rm s})\; \br{1-\frac{\chi}{\chi_{\rm s}}}\;,
}
where $p^{(i)}(\chi)$ denotes the probability distribution of comoving distances for a galaxy sample $i$. We choose to investigate a cross-power between two redshift bins of width 0.1, centred on $z=0.4$ and $z=0.8$, respectively. This signal is affected by correlations between the intrinsic sizes (or their FP size residuals) with the local matter distribution, which in turn contributes to the lensing effect on the background galaxies. This matches closely what we have measured via the $w_{{\rm g}\lambda}$ statistic. The corresponding systematic signal then reads \citep[see also][]{alsing15}
\eq{
\label{eq:limber_intrinsic}
C^{(ij)}_{\lambda \kappa}(\ell) = B \int^{\chi_{\rm hor}}_0 \dd
\chi\; \frac{p^{(i)}(\chi)\; q^{(j)}(\chi)}{\chi^2}\; P_{\delta
  \delta} \br{\frac{\ell}{\chi},\chi}\;,
}
where Equation (\ref{eq:lambdamodel}) was used once more.

\begin{figure}
\centering
\includegraphics[scale=.45]{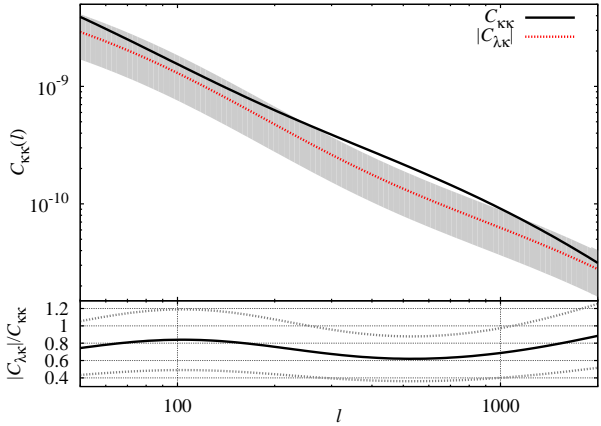}
\caption{Predicted contamination by intrinsic fundamental plane residual correlations of the
 tomographic convergence power spectrum,
  $C_{\kappa \kappa}(\ell)$, for the cross-power between two redshift bins of width 0.1
  centred on $z=0.4$ and $z=0.8$. In the top panel the cosmological signal is
  given by the black solid line and the absolute value of the systematic contribution by
  the red dotted line. The best-fit value for $B$ from the $w_{\rm g
    \lambda}$ fit ($r$-band, S13 sample as density tracer) has been propagated. Its propagated $1\sigma$ error results in the error envelope
  shown in grey. The bottom panel gives the ratio of the systematic over cosmological signal, with $1\sigma$ uncertainty indicated by the grey dotted lines.}
\label{fig:forecasts_mag}
\end{figure} 

Employing again the best-fit value and errors from the $r$-band fit to $w_{{\rm g}\lambda}$, using the S13 sample as density tracer, we obtain the power spectra shown in Fig.$\,$\ref{fig:forecasts_mag}. On all angular frequency scales that are accessible to measurement and modelling we find potentially severe contamination of the same order as the cosmological signal. Invoking the analogy with intrinsic alignments, this is perhaps not surprising because bright elliptical galaxies as found in the S13 sample also show the strongest alignment signals which would hamper a weak lensing shear measurement at these redshifts as well. It should be kept in mind though that, in order to make this prediction, we have used a simplistic model, extrapolated it substantially in redshift, and ignored a likely dependence on galaxy mass or luminosity. Clearly, however, intrinsic correlations of FP residuals need to be investigated in more detail before magnification measurements based on the FP become a viable cosmological probe. With a reasonable model for this systematic, weak lensing statistics in principle have the power to self-calibrate the different intrinsic correlations, as demonstrated by \citet{alsing15}.

\section{Conclusions}
\label{sec:conclusions}

Using a large sample of elliptical galaxies (defined as in \citealt{saulder13}) in the SDSS Main galaxy
sample, we report a $3-4\sigma$ detection of the spatial correlation function of fundamental plane (FP)
size residuals with the large-scale density field.  This detection is robust to the passband used to
make the flux and size measurements ($r$ or $i$) and to the choice of density tracers (the
elliptical galaxy sample or the entire Main galaxy sample).  A null detection of several systematics
signals, such as correlations at large separations along the line-of-sight or with shuffled sky
positions for the galaxies, rules out measurement-related systematics for our detection.

For further insight into our findings, we note that when splitting up the galaxies based on their
membership in groups, there is a clear trend for group central galaxies to have a positive residual
(meaning their observed size exceeds what is expected from the FP fits), and for
satellites to have a consistently negative size residual. We argue that these dependencies are likely to drive the negative correlation between matter density and FP residuals that we observe. Consistent with these trends, the clustering two-point correlation function of galaxies with positive FP residuals is lower than that for those with negative FP residuals, at $\sim 4\sigma$ significance.

Our findings can be interpreted as being caused by an above-average mass-to-light ratio for central galaxies and a below-average mass-to-light ratio for satellites, which is consistent with the standard picture for tidal stripping of satellites. In the group regime that we are mostly sensitive to there is multiple evidence in simulations for the stripping of dark matter haloes \citep[e.g.][]{kravtsov04} as well as the hot gaseous component of galaxies \citep{kawata08} during the infall into larger haloes. Observational results also point towards the existence of this effect; see for instance the direct measurements by \citet{suyu10} using strong lensing and Sif{\'o}n et al. (in prep.) using weak lensing. Clearly, a more quantitative comparison between the effects of tidal stripping and the spatial correlations in the FP residuals is desirable, but we leave this to forthcoming work.

We emphasise that the classic three-parameter FP as derived by \citet{saulder13} may be useful
to study the global properties of the galaxy sample, but cannot directly be employed in any cosmological application, nor in the work done in this paper. One reason is that the sample definition
with $1/V_\text{max}$ weighting is meant to remove the impact of the flux limit by upweighting the
lower luminosity galaxies that appear in the sample at low redshift but not higher redshift, so that
the {\em overall} sample luminosity distribution and derived FP is not subject to Malmquist bias.
However, when calculating correlation functions of FP residuals, or quantities derived therefrom, one always pairs up galaxies that are
at nearly the same redshift, and thus it is critical that the residuals be zero at {\em all
  redshifts}, not just on average, in order to avoid finding a spurious positive correlation
function. The $V_\text{max}$ weights conspire with an implicit lower limit on galaxy size, induced by the flux limit, to create a strong trend of increasing FP residuals as a function of redshift. We correct for these selection effects by explicitly including redshift as a fourth parameter in the FP fit.

Our detection has major implications for two types of cosmological measurements.  First, lensing
magnification studies that rely on galaxy size residuals as measured from the FP may be affected at the tens of percent level. Within our relatively large error bars we even cannot rule out that the intrinsic correlation attains the same amplitude as the cosmological weak lensing signal over a wide range of scales. \citet{ciarlariello14} explored a halo model in which galaxy intrinsic size correlates with the density field, which, however, is not directly comparable to our measurements because velocity dispersion and surface brightness variations also contribute to the FP residual correlations. Interestingly, they reached similar conclusions for the level of contamination expected for the size-convergence power spectrum as we did in Fig.~\ref{fig:forecasts_mag}. 

It will be elucidating to explore if, and to what extent, the correlations we detect can be attributed to effects of galaxy size, and to attempt a direct measurement of galaxy size intrinsic correlations, although the much increased intrinsic size noise contribution will make this challenging. Although it may appear that the predicted levels of astrophysical systematic could hinder the cosmological exploitation of size magnification, \citet{alsing15} showed that a joint analysis of shear and magnification of galaxy sizes can lead to substantial improvements in the
constraints on dark energy, of order $50\,\%$, even when simultaneously marginalising over flexible models for intrinsic galaxy alignments and size correlations. Studies analogous to galaxy-galaxy lensing, as conducted by \citet{huff14}, are not affected by this systematic as long as the lens and source galaxy samples are well separated (see \citealp{blazek12} for the equivalent study using galaxy ellipticities). Besides, the level of intrinsic correlations of residuals in the photometric analogue of the FP, as employed in that work, is currently unknown.

Second, attempts to use FP residuals as a distance indicator for the purpose of peculiar velocity studies could experience systematics due to intrinsic correlations at the $10\,\%$ level or above for wavenumbers $k \gtrsim 0.04\,h/$Mpc in the peculiar velocity auto-power spectrum.  Our $3$--$4\sigma$ detection does not give strong constraining power on the amplitude of the effect, and our ad-hoc model choice may induce further uncertainty in the forecast. Therefore, both observational and theoretical effort is warranted to get a better understanding of spatial FP residual correlations and their impact on peculiar velocity cosmography. Fortunately, one can identify several mitigation approaches for this systematic, including the dependence on the line-of-sight angle $\mu$ (since the intrinsic correlations are in projection on the sky), the use of self-calibration by incorporating cross-correlations between the galaxy distribution and peculiar velocities, as well as the comparison with samples relying on other distance indicators.

These findings underline that, despite past claims to the contrary, cosmology using galaxy surveys with today\rq s accuracy requirements is impossible without a thorough understanding of the underlying galaxy samples and their physical characteristics. The simultaneous inference on cosmological parameters and astrophysical effects related to intrinsic galaxy properties, such as the correlations found in this work, is therefore likely to be the default mode for the analysis of the forthcoming large spectroscopic and imaging surveys.

\section*{Acknowledgments}

We would like to thank Chris Blake and Eric Huff for stimulating discussions. We are grateful to our referee for an encouraging report. BJ acknowledges support by an STFC Ernest Rutherford Fellowship, grant reference ST/J004421/1. SS and RM acknowledge the support of the Department of Energy Early Career Award program.

Funding for SDSS-III has been provided by the Alfred P. Sloan
Foundation, the Participating Institutions, the National Science
Foundation, and the U.S. Department of Energy Office of Science. The
SDSS-III web site is http://www.sdss3.org/. SDSS-III is managed by the
Astrophysical Research Consortium for the Participating Institutions
of the SDSS-III Collaboration including the University of Arizona, the
Brazilian Participation Group, Brookhaven National Laboratory,
Carnegie Mellon University, University of Florida, the French
Participation Group, the German Participation Group, Harvard
University, the Instituto de Astrofisica de Canarias, the Michigan
State/Notre Dame/JINA Participation Group, Johns Hopkins University,
Lawrence Berkeley National Laboratory, Max Planck Institute for
Astrophysics, Max Planck Institute for Extraterrestrial Physics, New
Mexico State University, New York University, Ohio State University,
Pennsylvania State University, University of Portsmouth, Princeton
University, the Spanish Participation Group, University of Tokyo,
University of Utah, Vanderbilt University, University of Virginia,
University of Washington, and Yale University. 

\bibliographystyle{mn2e2}%sukhdeep: I've added bst file mn2e2, which overcomes issues with arxiv and long author lists
\bibliography{bibliography}

\label{lastpage}
\end{document}